\documentclass[lettersize,journal]{IEEEtran}

\usepackage{textcomp}
\usepackage{url}
\usepackage{verbatim}
\usepackage{mathrsfs}
\usepackage{mathtools}
\usepackage{dsfont}
\usepackage{relsize}
\usepackage{graphicx}
\usepackage{amsmath}
\usepackage{lipsum}
\usepackage{algcompatible}
\usepackage{algpseudocode}
\usepackage{amsthm}
\usepackage{amsfonts}
\usepackage{caption}
\usepackage{blkarray}
\usepackage{mathtools}
\usepackage{siunitx}
\usepackage{amssymb}
\usepackage{subcaption}
\usepackage{placeins}
\usepackage{xcolor}
\usepackage{lettrine}
\usepackage{multicol, blindtext}
\usepackage{bbm}
\usepackage{cite}
\usepackage{algorithm}
\usepackage{hyperref}
\usepackage{algpseudocode}
\usepackage{enumitem}

\newtheorem{remark}{Remark}

\makeatletter

\newcommand{\Rmnum}[1]{\expandafter\@slowromancap\romannumeral #1@}

\makeatletter
\let\NAT@parse\undefined
\makeatother
\usepackage{hyperref}  
\hypersetup{
    colorlinks= true,
    citecolor= black,
    linkcolor= black, 
    pdfborder= {0 0 0},
}

\newtheorem{theorem}{Theorem}

\newtheorem{proposition}{Proposition}

\newtheorem{definition}{Definition}
\newtheorem{lemma}{Lemma}

\makeatletter
\def\BState{\State\hskip-\ALG@thistlm}
\makeatother

\def\BibTeX{{\rm B\kern-.05em{\sc i\kern-.025em b}\kern-.08em
    T\kern-.1667em\lower.7ex\hbox{E}\kern-.125emX}}

\begin{document}
\title{Optimal Denial-of-Service Attacks Against\\Partially-Observable Real-Time Monitoring Systems}

\makeatletter
\author{
\IEEEauthorblockN{Saad Kriouile, Mohamad Assaad, Amira Alloum, and Touraj Soleymani}
\thanks{Saad Kriouile (saad.kriouile@centralesupelec.fr) and Mohamad Assaad (Mohamad.Assaad@centralesupelec.fr) are with the Laboratory of Signals and Systems, CentraleSup\'{e}lec, University of Paris-Saclay, 91190 Gif-sur-Yvette, France. Amira Alloum is with Qualcomm, France (aalloum@qti.qualcomm.com). Touraj Soleymani is with the City St George's School of Science and Technology, University of London, London EC1V~0HB, United Kingdom (touraj@city.ac.uk). This work has received funding from Qualcomm.}%
}


\maketitle
\begin{abstract}
In this paper, we study the impact of denial-of-service attacks on the status updating of a cyber-physical system with one or more sensors connected to a remote monitor via unreliable channels in partially-observable environment. We approach the problem from the perspective of an adversary that can strategically jam a subset of the channels. The sources are modeled as Markov chains, and the performance of status updating is measured based on the age of incorrect information at the monitor. Our goal is to derive jamming policies that severely degrade the system's performance while operating within the adversary's energy constraint. For a single-source scenario, we formulate the problem as a partially-observable Markov decision process, and rigorously prove that the optimal jamming policy is of a threshold form. We then extend the problem to a multi-source scenario. We formulate this problem as a restless multi-armed bandit, and provide a jamming policy based on the Whittle's index. Our numerical results highlight the performance of our policies compared to baseline policies.
\end{abstract}

\begin{IEEEkeywords}
age of incorrect information, cyber attacks, cyber-physical systems, jamming policies, networks, status updating, real-time systems.
\end{IEEEkeywords}

\section{Introduction}

Cyber-physical systems (CPSs) represent a sophisticated fusion of computational algorithms and physical processes, bridging the gap between the digital and real-world environments~\cite{kim2012cyber}. These systems leverage advanced computation, communication, and control mechanisms to enhance performance, adaptability, and automation across various sectors. The dynamic nature of CPSs demands continuous real-time monitoring to capture the most recent changes in their operating environments. Such a stream of up-to-date data allows CPSs to swiftly adapt to evolving conditions, ensuring decisions are based on accurate and timely information~\cite{voi, voi2, soleymani2024retransmit, maatouk2022age, kriouile2021global, sun2017}. Nevertheless, CPSs are inherently susceptible to a wide range of cyber threats and attacks~\cite{humayed2017cyber}. Three prevalent types of cyber attacks include deception attacks, eavesdropping attacks, and denial-of-service attacks, as noted in~\cite{cardenas2008secure}. Deception attacks compromise data integrity, eavesdropping attacks breach confidentiality, and denial-of-service attacks---the focus of our study---disrupt data availability by obstructing network transmissions. We here investigate the impact of denial-of-service attacks on the status updating of a CPS with one or more sensors connected to a remote monitor via unreliable channels. We approach this problem from the perspective of an adversary that can strategically jam the channels. Our objective is to derive jamming policies that severely degrade the system's performance while operating within the adversary's energy constraint. Determining such policies enable us to gain insight into the potential damage that such attacks can inflict on monitoring~systems.

In our study, the sources are modeled as Markov chains, and the performance of status updating is measured based on the age of incorrect information (AoII) at the monitor in a partially-observable environment. The AoII was initially proposed in~\cite{maatouk2020age}, which takes into account both the change of the source status and the freshness of information. The reader is referred to~\cite{kriouile2021aoii, maatouk2022age, kam2020age, yates2021age, uysal2022semantic}, and the references therein, for more details about this semantic metric and its use for CPSs. Notably, the works in~\cite{maatouk2020age,maatouk2022age,kam2020age} developed optimal scheduling policies that minimize the AoII in real-time monitoring applications, where it is assumed that the scheduler has a perfect knowledge of the status of the source at each time. However, in practice, specially if multiple sources need to be tracked, the scheduler might not be aware of the status of the sources~\cite{kriouile2021aoii}. In this partially-observable environment, the AoII is not known, and must be estimated.

Previous research have explored the designing jamming policies within the context of networked estimation and networked control~\cite{zhang2015optimal, zhang2015optimalenergy, qin2017optimal, zhang2016attackengergy, qin2020optimal, gan2020optimizing, zhang2018denial}. For instance, optimal jamming policy maximizing a regulation loss subject to a frequency constraint has been developed in  \cite{zhang2015optimal}.  Packet-drop-ratio stealthiness is considered in \cite{zhang2015optimalenergy} where a distortion loss subject to a frequency constraint is optimized using signal-independent jamming policy. \cite{qin2017optimal} considers a channel with inherent packet loss for multi-source scenario and identified an optimal signal-independent jamming policy that maximizes distortion loss subject to a frequency constraint. In \cite{zhang2016attackengergy},  static and dynamic attack scenarios are considered and optimal jamming policy that maximizes distortion loss subject to a power constraint is developed, while considering that the channel is ideal in the absence of attacks. Building on~\cite{zhang2016attackengergy}, a more challenging scenario is considered in \cite{qin2020optimal} where packet loss can occur even without any attacks. In \cite{gan2020optimizing}, a scenario where sensors communicate with monitors via two-hop relay channels is considered, and jamming policies that maximize distortion subject a power constraint are explored. Lastly, a regulation loss maximization subject to a soft power constraint in a fading channel is considered, where an optimal jamming policy is  developed and the impact of various packet detection~schemes is analyzed.

Only a limited number of studies have explored the design of jamming policies using freshness metrics~\cite{nguyen2017jamming, xiao2018jamming, yang2022jamming, kriouile2024optimal}. Notably, Nguyen~\emph{et~al.}~\cite{nguyen2017jamming} framed the interaction between an adversary and a scheduler as a static game by utilizing the age of information (AoI) as the performance index, and derived the optimal jamming power level for a channel represented by an $M$/$G$/$1$/$1$ queue. Xiao~\emph{et~al.}\cite{xiao2018jamming} formulated such an interaction as a dynamic game by utilizing the AoI as the performance index, and derived the optimal jamming time distribution. Yang~\emph{et~al.}\cite{yang2022jamming} studied the distributed coordination of multiple sensors’ channel access behavior under jamming attacks by utilizing an AoI-based performance index, and derived optimal channel access policies when the channels are modeled as $M$/$M$/$1$, $M$/$D$/$1$, or $D$/$M$/$1$ queues. Finally, Kriouile~\emph{et~al.}~\cite{kriouile2024optimal} considered a single-source scenario by utilizing basic AoI and AoII-based performance indices, and derived the corresponding optimal jamming policies of an omniscient adversary that knows the mismatch between the true state and the estimated state at the monitor at each~time. Unlike the work in~\cite{kriouile2024optimal}, we here consider an environment where the adversary does not have exact knowledge of the source states at each time. In addition, we address a multi-source scenario where the adversary selectively jams a subset of the channels.

\subsection{Overview and Organization}
In this paper, we propose a novel framework for designing jamming policies that can degrade the performance of monitoring systems, thereby impairing their abilities to accurately track and respond to real-time events. We seek to find jamming policies in single and multi-source scenarios. For a single-source scenario, we formulate the problem as a \emph{partially-observable Markov decision process} (POMDP), and rigorously prove that the optimal jamming policy is of a threshold form. We then extend the problem to a multi-source scenario. We formulate this problem as a \emph{restless multi-armed bandit} (RMAB), and provide a jamming policy based on the Whittle's index. Our numerical results highlight the performance of our policies compared to baseline policies. We should emphasize that our study departs from the previous works in the literature. In particular, in contrast to~\cite{nguyen2017jamming, xiao2018jamming, yang2022jamming}, we here consider a utility function based on the AoII. In addition, on the contrary to~\cite{kriouile2024optimal}, we here consider more realistic single-source and multi-source scenarios where the adversary does not know exactly the states of the sources at each time.

The paper is organized as follows. We formulate the problems mathematically and present our main theoretical results for the single-source and multi-source scenarios in Sections~\ref{sec:single-source} and \ref{sec:multi-source}, respectively. We provide our numerical results in Section~\ref{sec:num_reslt}. Finally, we conclude the paper and discuss potential future research directions in Section~\ref{sec:conclusion}.

\section{Single-Source Scenario}\label{sec:single-source}
In this section, we study the single-source scenario in which we deal with a single source that sends its status updates to a remote monitor over an unreliable channel, and an adversary that can strategically jam the channel. First, we present the system model, and define the main parameters of the system. Then, we describe the evolution of the AoII in a partially-observable environment. Finally, we formulate the problem as a POMDP, and derive the optimal jamming policy.

\subsection{Networked System Model}\label{subsec:Net_descrip}
Our system model consists of a binary Markovian source, a remote monitor/receiver and an adversary. Time is discretized into uniform time slots. At each time slot, the Markovian source generates an update packet that is sent instantaneously over an unreliable channel to the monitor, contingent upon the state of the source. The channel's quality, denoted as $c(t)$, is modeled as an independent and identically distributed (iid) process with two possible states: $0$ and $1$. When $c(t) = 1$, the packet is successfully decoded by the receiver, and when $c(t) = 0$, the packet fails. The probability of successful decoding, $c(t) = 1$, is given by $p$. Upon successful reception, the monitor sends an acknowledgment (ACK) back to the transmitter; otherwise, it sends a negative acknowledgment (NACK).
For the Markovian source, the state transition probability, denoted by $r$, is the probability of transitioning from one state to another. We assume $r \leq 1/2$, meaning the source is more likely to remain in its current state than to switch. The adversary, acting as malicious agent within our system model, attempts to jam the channel. Let $d(t)$ denote the action of the adversary at time $t$ such that $d(t)=1$ if the adversary attempts to jam the channel, and $d(t)=0$ otherwise. When the adversary decides to jam the channel, the probability that it succeeds is equal to $q$. We consider that the adversary can intercept the ACK and NACK messages at~each~time.

We consider that a positive cost (a negative reward) is incurred when the adversary attempts to jam the channel due
to the energy required for jamming.

\subsection{Performance Index and It's Dynamics}
In our study, the performance of status updating is measured based on the AoII at the monitor. If the source sends a packet only when the estimated state at the monitor differs from that of the source, the adversary will be able to track the mismatch between the true state and the estimated state (see \cite{kriouile2024optimal} for more details). In particular, if the monitor does not send any ACK or NACK acknowledgement, the adversary infers that the true state and the estimated state are the same; otherwise, it infers that they are different. Therefore, from the source's perspective, to further obscure the adversary in tracking the mismatch, it is more appropriate if the source sends a packet at every time. In this way, the adversary will not be able to discern the mismatch as long as the monitor fails to decode the received packet. Consequently, the adversary should estimate the probability of the mismatch, and compute the expected age of incorrect information (EAoII).

Let $X(t)$, $\hat{X}(t)$, and $g(t)$ be the source's state, the estimated state at the monitor, and the time stamp of the last successfully received packet by the monitor, respectively. When the monitor does not receive any new status update, it uses the last successfully received packet as the estimated state. Hence, $\hat{X}(t)= X(g(t))$. Accordingly, the explicit expression of the EAoII can be written as
\begin{align}
s(t)=\mathbb{E}\Big[(t-V(t)\Big]
\end{align}
where $s(t)$ represents the value of the EAoII at time $t$, and $V(t)$ is the last time instant such that
\begin{align}
X(V(t))= \hat{X}(g(t)).
\end{align}
Now, by \cite[Lemmas 1 and 2]{kriouile2021aoii}, we can establish that
\begin{align} \label{eq:exp_aoii}
s(t) &=\mathbb{E} \Big[t-V(t) \Big] \nonumber \\
&=\frac{1}{2r} \Big[1+(1-2r)^{t-g(t)+1}-2(1-r)^{t-g(t)+1} \Big] .
\end{align}
According to \eqref{eq:exp_aoii}, we have $s(t) = s_{t-g(t)}$, where $s_k=\frac{1}{2r} [1+(1-2r)^{k+1}-2(1-r)^{k+1}]$ for all $k \ge 0$.

Note that, at time $t+1$, if the packet is successively transmitted, then $g(t+1)=t+1$. Therefore, at time $t+1$, the EAoII equals to $s(t+1)=s_{t+1-g(t+1)}=s_0$. This occurs if $c(t+1)=1$ and $d(t)=0$; or $c(t+1)=1$ and $d(t)=1$ but the adversary does not succeed in jamming the channel. However, if the packet is not successively transmitted, then $g(t+1)=g(t)$. Therefore, the EAoII will transition to $s(t+1)=s_{t+1-g(t+1)}=s_{t-g(t)+1}$. Following this observation, the transition probabilities for the EAoII are obtained as
\begin{align} \label{eq:trans_pb}
    &\Pr\big(s(t+1)=s_0|s(t)=s_k, d(t)=0\big)=p, \nonumber \\[1.5\jot]
    &\Pr\big(s(t+1)=s_{k+1}|s(t)=s_k, d(t)=0\big)=1-p, \nonumber\\[1.5\jot]
    &\Pr\big(s(t+1)=s_0|s(t)=s_k, d(t)=1\big)=p(1-q), \nonumber \\[1.5\jot]
    &\Pr\big(s(t+1)=s_{k+1}|s(t)=s_k, d(t)=1\big)=1\!- p(1\!-\!q) .
\end{align}

\subsection{Problem Statement}
In this paper, we aim to determine the optimal jamming policy that maximizes the long-term expected average reward. Specifically, we seek to identify the sequence of actions, denoted as $\phi = (d^{\phi}(0), d^{\phi}(1), \dots)$, where $d^{\phi}(t)$ represents the jamming action taken by the attacker at time slot $t$. In this framework, $d^{\phi}(t) = 1$ indicates that the adversary chooses to jam the channel at time $t$, and $d^{\phi}(t) = 0$ otherwise. Taking into account the energy cost associated with jamming, the adversary’s reward at time $t$ is defined as $R(t) = s(t) - \lambda d(t)$, where $\lambda > 0$ represents the jamming cost. Our objective is to solve the following optimization problem
\begin{align}\label{eq:problem_formulation}
& \underset{\phi\in \Phi}{\text{maximize}} \ \lim_{T\to\infty} \text{inf}\:\frac{1}{T}\mathbb{E}^{\phi\in \Phi}\Big[ \sum_{t=0}^{T-1}R^{\phi}(t) \big|s(0)\Big]
\end{align}
where $\Phi$ denotes the set of all causal jamming policies and $R^{\phi}(t)$ denotes the reward function under the jamming policy~$\phi$.

\subsection{Optimal Jamming Policy}\label{sec:main_results}
In the rest of this section, we provide our theoretical results on the design of the optimal jamming policy. We begin by showing that the optimal solution is a threshold policy. Then, we provide a closed-form expression of the problem of interest, and derive the optimal threshold value as a function~of~$\lambda$.
We reformulate our problem in (\ref{eq:problem_formulation}) as infinite-horizon average-reward partially Observable Markov Decision Problem (POMDP) where $s(t)$ is the observed state at time $t$ referring to the value of EAoII, $d(t)$ the jamming action determining if the channel is under attack or not, $Pr(s \rightarrow s'|d)$ the transition probabilities from state $s$ to $s'$ under action $d$ that satisfy the equations given in \eqref{eq:trans_pb} and $R(t)=s(t)-\lambda d(t)$ the instantaneous reward incurred at time $t$.

The optimal policy denoted $\phi^*$ of the problem in~\eqref{eq:problem_formulation} is the one that solves the following Bellman optimality equation for each state $s = s(t)$:
\begin{align}
\theta + V(s) =\max_{d\in\{0,1\}} \Big\{s-\lambda d+\sum_{s'\in \mathbb{S} }\Pr(s\rightarrow s'|d)V(s') \Big\} 
\label{eq:bellman_general}
\end{align}
where $\theta$ is the optimal value of the problem, $V(s)$ is the differential reward-to-go function, and $\mathbb{S}$ is a set defined as
\begin{align}
\mathbb{S} = \Big \{ \frac{1}{2r}\big [1+(1-2r)^{k+1}-2(1-r)^{k+1} \big ] \big| k \ge 0 \Big \}.
\end{align}

In the next two lemmas, we show that $s_i$ and $V(s_i)$ are increasing functions.
\begin{lemma}\label{lem:aoui_increasing}
$s_i$ is an increasing function with respect to $i$
\end{lemma}
\begin{IEEEproof}
We have $s_{i+1}-s_i=(1-r)^{i+1}-(1-2r)^{i+1}$. Knowing that $r \le 1/2$, for all integer $i$, we have $(1-r)^{i+1}-(1-2r)^{i+1} \ge 0$. That means that $s_{i+1}-s_i \ge 0$.   
\end{IEEEproof}

\begin{lemma}\label{lem:V_increasing}
$V(s_i)$ is an increasing function with respect to $s_i$.
\end{lemma}
\begin{IEEEproof}
The relative value iteration equation consists of updating the value function $V^t(.)$ as follows: 
\begin{align}
V_{t+1}(s_i)=\max\big\{V_t^0(s_i),V_t^1(s_i)\big\} 
\end{align}
where
\begin{align}
V_t^0(s_i)& = s_i+(1-p)V_t(s_{i+1})+pV_t(s_0), \\[1.5\jot]
V_t^1(s_i)& = s_i-\lambda+(pq+1-p)V_t(s_{i+1}) \nonumber\\[1.5\jot]
            &\qquad +(1-q)pV_t(s_0) .
\end{align}
Our proof is by induction. We show that $V_t(\cdot)$ is increasing for all $t$ and we conclude that for $V(\cdot)$. As $V_0(.)=0$, then the property holds for $t=0$. If $V_t(.)$ is increasing with respect to $s_i$, we show that, for $s_i\le s_j$, we have $V_{t+1}^0(s_i) \leq V_{t+1}^0(s_{j})$ and $V_{t+1}^1(s_i) \leq V_{t+1}^1(s_j)$.

We can write
\begin{align}
&V_{t+1}^0(s_{j})-V_{t+1}^0(s_i) \nonumber\\[1.5\jot]
&\qquad \quad = s_j-s_i+(1-p)(V_{t}(s_{j+1}) - V_{t}(s_{i+1})).
\end{align}
By Lemma \ref{lem:aoui_increasing}, we have $s_{i+1} \le s_{j+1}$. Hence, since $V_t(.)$ is increasing with respect to $s_i$, therefore $V_{t+1}^0(s_{j}) - V_{t+1}^0(s_i) \geq 0$. As consequence, $V_{t+1}^0(\cdot)$ is increasing with respect to $s_i$. 

Similarly, we can write
\begin{align}
&V_{t+1}^1(s_{j}) - V_{t+1}^1(s_{i})\nonumber\\[1.5\jot]
& \quad =s_j-s_i + (1-p+pq) (V_{t}(s_{j+1}) - V_{t}(s_{i+1})).
\end{align}
Hence, $V_{t+1}^1(s_j) - V_{t+1}^1(s_i) \geq 0$. As consequence, $V_{t+1}^1(\cdot)$ is increasing with respect to $s_i$.

Now, since $V_{t+1}(.)=\max\{V^0_{t+1}(\cdot),V^1_{t+1}(\cdot)\}$, then $V_{t+1}(.)$ is increasing with respect to $s_i$. We demonstrated by induction that $V_t(.)$ is increasing for all $t$. Knowing that $\lim_{t \to \infty} V_t(s_i)=V(s_i)$, we conclude that $V(.)$ is also increasing with respect to $s_i$.
\end{IEEEproof}

In the next theorem, we specify the structure of the optimal jamming policy.

\begin{theorem}\label{theo:threshold_policy}
The optimal jamming policy associated with the problem in (\ref{eq:bellman_general}) and for any given $\lambda$ is an increasing threshold policy. More specifically, there exists $n \in \mathbb{N}$ such that the prescribed action is passive, i.e., $d_k = 0$, when $s_k < s_n$, and the prescribed action is active, i.e., $d_k = 1$, when $s_k \geq s_n$.
\end{theorem}
\begin{IEEEproof}
Note that the explicit expression of the Bellman optimality equation when we are in state $s_i$ is written as
\begin{align}
&\theta + V(s_i) = \max \Big\{ s_i + pV(s_0)+(1-p)V(s_{i+1}); \nonumber\\[1.5\jot]
&\qquad \qquad s_i-\lambda+(q+(1-p)(1-q)) V(s_{i+1}) \nonumber\\[1.5\jot]
&\qquad \qquad +(1-q)p V(s_0) \Big\} .
\end{align}
We will deduce that the optimal solution is a threshold-increasing policy by establishing that $\Delta V(s)=V^1(s_i)-V^0(s_i)$ is increasing in $s_i$, where $V^1(s_i)$ and $V^0(s_i)$ are the value functions evaluated at $s_i$, when the action is equal to $1$ and $0$, respectively. More specifically, we can write
\begin{equation}
\Delta V(s_i)=V^1(s_i)-V^0(s_i)
\end{equation}
where $\lim_{t \to \infty} V_t^0(s_i)=V^0(s_i)$ and $\lim_{t \to \infty} V_t^1(s_i)=V^1(s_i)$. Subsequently, $\Delta V(s_i)$ equals to:
\begin{equation}
\Delta V(s_i)=-\lambda+pq(V(s_{i+1})-V(s_0)).
\end{equation}
According to Lemma~\ref{lem:V_increasing}, $V(.)$ is increasing with respect to $s_i$. Therefore, $\Delta V(s_i)$ is also increasing with $s_i$. Moreover, by Lemma~\ref{lem:aoui_increasing}, $s_i$ is increasing with respect to $i$. Hence, there exists $n \in \mathbb{N}$ such that $\Delta V(s_i)\leq 0$ for all $s_i < s_n$, and $\Delta V(s_i) \ge 0$ for all $s_i \ge s_n$. Given that the optimal action at state $s_i$ is the one that maximizes $\{V^0(\cdot),V^1(\cdot)\}$, then, for all $s_i < s_n$, the optimal decision is the passive one since $\max\{V^0(s_i),V^1(s_i)\}=V^0(s_i)$, and, for all $s_i > s_n$, the optimal decision is the active one since $\max\{V^0(s_i),V^1(s_i)\}=V^1(s_i)$. This concludes the proof.
\end{IEEEproof}

\begin{remark}
According to Theorem~\ref{theo:threshold_policy}, when the EAoII is less than the optimal threshold, the impact of the adversary's action on the reward is small, whereas this impact becomes more important when the EAoII is larger than the optimal threshold. Consequently, the adversary should save energy until the EAoII becomes large enough to reach the optimal threshold, which happens inevitably due to the existence of an unreliable channel, and then should jam the channel and keep doing this as far as the EAoII is larger than the optimal threshold. Note that this result is quite important as it dramatically simplifies the structure of the jamming policy, which is advantageous in practice.
\end{remark}

\begin{figure}[t] 
\centering
\includegraphics[width=0.5\textwidth]{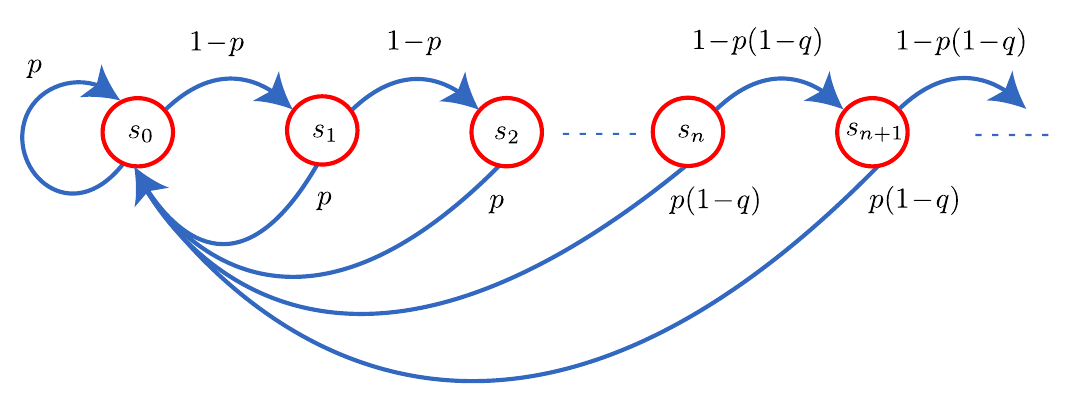}
\caption{The state transition under a threshold jamming policy with parameter $s_n$, where the state here is the EAoII.}
\label{fig:dtmc}
\end{figure}

\subsection{Closed-Form Expression of the Optimal Solution}
In this section,  we find the explicit expression of the optimal threshold value in function of the systems parameters and $\lambda$. For that, we introduce the steady-state from of the the problem \eqref{eq:problem_formulation} as follows:
\begin{align}\label{eq:closed_form_express}
& \underset{n\in \mathbb{N}}{\text{maximize}} \ \overline{s}_{n}-\lambda\overline{d}_n
\end{align}
where $\overline{s}_{n}$ is the average EAoII, and $\overline{d}_n$ is the average active attacking time (AAT) under the threshold jamming policy with parameter~$s_n$. More specifically, we have 
\begin{align}
\overline{s}_{n}&=\lim_{T\to\infty} \text{inf}\:\frac{1}{T}\mathbb{E}^{n}\Big[\sum_{t=0}^{T-1}s(t) \big|s(0),\phi^t(n)\Big],\label{eq:average_age}\\
\overline{d}_n&=\lim_{T\to\infty} \text{inf}\:\frac{1}{T}\mathbb{E}^{n}\Big[\sum_{t=0}^{T-1}d(t) \big|s(0),\phi^t(n)\Big]\label{eq:average_active_time}
\end{align}
where $\phi^t(n)$ denotes the threshold jamming policy with parameter $s_n$.

To get the expressions of $\overline{s}_{n}$ and $\overline{d}_n$, we will compute the stationary distribution of the discrete-time Markov chain (DTMC) that models the evolution of the EAoII under the threshold policy with parameter $s_n$, as illustrated in Fig.~\ref{fig:dtmc}.

\begin{proposition}\label{prop:stationary_distribution}
For a given threshold jamming policy with parameter $s_n$, the DTMC admits $u_n(s_i)$ as its stationary distribution:
\begin{align}\label{stationarydistribution}
 u_n(s_i)=\left\{
    \begin{array}{ll}
        \frac{(1-q)p}{1-q+q(1-p)^{n}} & \text{if} \ i=0,  \\[3\jot]
        (1\!-\!p)^{i}\frac{(1-q)p}{1-q+q(1-p)^{n}} & \text{if} \ 1 \leq i \leq n,  \\[3\jot]
        (1\!-\!p)^n(pq+1\!-\!p)^{i-n} & \text{if} \ i \geq n+1 .\\[1\jot]
        \qquad \qquad \times\frac{(1-q)p}{1-q+q(1-p)^{n}} 
    \end{array}
\right.
\end{align}
\end{proposition}
\begin{IEEEproof}
In order to find the expression of the stationary distribution, we should resolve the following full balance equation:
\begin{align}
    u_n(s_k)=\sum_{j=0}^{\infty} {\Pr}^n(s_j \rightarrow s_k) u_n(s_j)
\end{align}
where $\Pr^n(s_j \rightarrow s_k)$ is the transition probability from $s_j$ to $s_k$ under the threshold jamming policy with parameter $s_n$. We begin with the expression of $\Pr^n(s_j \rightarrow s_k)$ for all $j$ and $k$. To that end, we can distinguish three cases of $k$.

\underline{Case I} ($k=0$): According to Theorem \ref{theo:threshold_policy}, if $s_j < s_n$, the adversary stays idle, thus $s_j$ moves to $s_0$ with probability $p$. Otherwise, it moves to $s_0$ with probability $p(1-q)$. Therefore, we get $\Pr^k(s_j \rightarrow s_0)=p$ if $s_j < s_n$; and $\Pr^k(s_j \rightarrow s_0)=p(1-q)$ if $s_j \ge s_n$.

\underline{Case II} ($0< k \le n$; $s_k \le s_n$ as $s_k$ is increasing with $k$): From (\ref{eq:trans_pb}), the only state that is susceptible to transition to $s_k$ at the next time is $s_{k-1}$ with probability $1-p$. Thus $\Pr^k(s_j \rightarrow s_k)=(1-p) \mathds{1}_{j=k-1}$.

\underline{Case III} ($k > n$): From (\ref{eq:trans_pb}), the only state that is susceptible to transition to $s_k$ at the next time is $s_{k-1}$ with probability $1-p(1-q)$. Thus $\Pr^n(s_j \rightarrow s_k)=(1-p(1-q))\mathds{1}_{j=k-1}$.

The above results imply that $u_n(s_k)=(1-p)u_n(s_{k-1})$ for $0<k \le n$; and $u_n(s_k)=p(1-q)u_n(s_{k-1})$ for $k >n$. Hence, we can show by induction that $u_n(s_k)=(1-p)^k u_n(s_{0})$ for $0<k \le n$; and $u_n(s_k)=(1-p)^n(1-p(1-q))^{i-n}u_n(s_0)$ for $k >n$. The last step is to derive $u_n(s_0)$. Following the fact that $\sum_{k=0}^{\infty}u_n(s_k)=1$, we can establish that
\begin{align*}
u_n(s_0)= \frac{(1-q)p}{1-q+q(1-p)^{n}}.
\end{align*}
This completes the proof.
\end{IEEEproof}

In the next two propositions, we derive closed-form expressions for the average EAoII and the average AAT under the threshold jamming policy with parameter $n$.
\begin{proposition}\label{prop:average_reward_expression}
Under a threshold jamming policy $s_n$, the average EAoII $\overline{s}_{n}$ is equal to
\begin{align}\label{eq:ss-EAoII}
\overline{s}_{n} &= \frac{(1-q)p}{1-q+q(1-p)^{n}} \bigg[\frac{r(1-p)}{p(1-y)(1-z)}\nonumber\\[2\jot]
&\quad +\frac{q(1-p)^n}{2pr(1-q)}-\frac{pq(1-p)^n(1-r)^{n+2}}{r(1-z)\big(1-x\big)}\nonumber \\[2\jot]
&\quad +\frac{pq(1-p)^n(1-2r)^{n+2}}{2r(1-y)\big(1-x\big)}\bigg]
\end{align}
where $y = (1-2r)(1-p)$, $z = (1-r)(1-p)$, $w=(1-p+pq)(1-r)$ and $x=(1-p+pq)(1-2r)$
\end{proposition}
\begin{IEEEproof}
Note that, in the steady state, $\overline{s}_n=\sum_{k=0}^{\infty} k u_n(s_k)$. Hence, exploiting the expression of $u_n(\cdot)$ given in Proposition~\ref{prop:stationary_distribution}, we obtain the result.
\end{IEEEproof}

\begin{proposition}\label{prop:average_active_time}
Under a threshold jamming policy $s_n$, the average AAT $\overline{d}_n$ is equal~to
\begin{align}\label{eq:ss-AAT}
\overline{d}_n=\frac{(1-p)^n}{(1-q)+q(1-p)^n} .
\end{align}
\end{proposition}
\begin{IEEEproof}
Again, in the steady state, $\overline{d}_n=\sum_{k=n}^{\infty} u_n(s_k)$. Hence, exploiting the expression of $u_n(\cdot)$ given in Proposition~\ref{prop:stationary_distribution}, we can obtain the result.
\end{IEEEproof}

Leveraging Propositions \ref{prop:average_reward_expression} and Proposition \ref{prop:average_active_time}, we define the sequence $\lambda(s_n)$ in the following definition, with its exact expression resembling that in~\eqref{eq:exp_lambda}.
\begin{definition}\label{def:intersection_poin_lambda}
$\lambda(s_n)$ is the intersection point between $\overline{s}_n-\lambda\overline{d}_n$ and $\overline{s}_{n+1}-\lambda \overline{d}_{n+1}$, i.e.,
\begin{align}
\lambda(s_n)=\frac{\overline{s}_{n+1}-\overline{s}_{n}}{\overline{d}_{n+1}-\overline{d}_{n}}.
\end{align}
\end{definition}

In the following two lemmas, we show a few important properties of the sequence $\lambda(s_n)$, which are instrumental for our~analysis.
\begin{lemma}\label{lem:increasing_whittle_index}
The sequence $\lambda(s_n)$ is strictly increasing with respect to $n$.
\end{lemma}
\begin{IEEEproof}
It is not difficult to show that
\begin{align}
    \lambda(s_{n+1})-\lambda(s_n)=&(1-q+(1-p)^{n+1}q)pq\nonumber \\[2\jot]
    &  \times \bigg[\frac{(1-r)^{n+2}}{1-(1-r)(1-p+pq)} \nonumber \\[2\jot]
    & \ \ \ \ -\frac{(1-2r)^{n+2}}{1-(1-2r)(1-p+pq)} \bigg].
\end{align}
We know that the function
\begin{align*}
f(x) = \frac{x^{n+2}}{1-x(1-p+pq)}
\end{align*}
is increasing with respect to $x$ for $x \ge 0$. Therefore, as $1-r \ge 1-2r \ge 0$, we have $\lambda(s_{n+1})-\lambda(s_n) \ge 0$. Consequently, we deduce that $\lambda(.)$ is increasing with respect to $n$. 
\end{IEEEproof}

\begin{lemma}
The sequence $\lambda(s_n)$ satisfies the following conditions for any given $\lambda$:
 \begin{align}
    &\overline{s}_k \!-\! \lambda \overline{d}_k \ge \overline{s}_{k+1} \!-\! \lambda \overline{d}_{k+1} \ \text{for} \ \lambda \leq \lambda(s_k), \label{eq:L1}\\[1.5\jot]
    &\overline{s}_k \!-\! \lambda \overline{d}_k < \overline{s}_{k+1} \!-\! \lambda \overline{d}_{k+1} \ \text{for} \ \lambda > \lambda(s_k), \label{eq:L2}\\[1.5\jot]
    &\overline{s}_{n+1} \!-\! \lambda \overline{d}_{n+1} > \overline{s}_{k} \!-\! \lambda \overline{d}_k \ \text{for} \ \lambda > \lambda(s_n), \ k \le n, \label{eq:L3}\\[1.5\jot]
    &\overline{s}_{n+1} \!-\! \lambda \overline{d}_{n+1} \geq \overline{s}_{k} \!-\! \lambda \overline{d}_k \ \text{for} \ \lambda \le \lambda(s_{n\!+\!1}), \ k > n. \label{eq:L4}
\end{align}   
\end{lemma}
\begin{IEEEproof}
By definition of $\lambda(s_k)$, $\overline{s}_k - \lambda(s_k) \overline{d}_k = \overline{s}_{k+1} - \lambda(s_k) \overline{d}_{k+1}$.
Therefore, we have that:
\begin{align*}
&(\overline{s}_k - \lambda \overline{d}_k) - (\overline{s}_{k+1} - \lambda \overline{d}_{k+1}) \nonumber \\[1.5\jot]
&\qquad = \lambda(s_k) \overline{d}_k- \lambda(s_k) \overline{d}_{k+1}-\lambda  \overline{d}_k+\lambda  \overline{d}_{k+1} \nonumber \\[1.5\jot]
&\qquad = (\lambda - \lambda(s_k)) (\overline{d}_{k+1} - \overline{d}_k).
\end{align*} 
Given that $\overline{d}_k$ is strictly decreasing with respect to $k$, we can write $\overline{s}_k - \lambda \overline{d}_k \ge \overline{s}_{k+1} - \lambda \overline{d}_{k+1}$ when $\lambda \leq \lambda(s_k)$, and $\overline{s}_k - \lambda \overline{d}_k < \overline{s}_{k+1} - \lambda \overline{d}_{k+1}$ when $\lambda > \lambda(s_k)$. This completes the proof of the first and second conditions.

In order to prove the third condition, it is sufficient to show that $\overline{s}_{k} - \lambda \overline{d}_k$ is strictly increasing with respect to $k$ when $k \leq n$ and $\lambda > \lambda(s_n)$. To that end, we prove that $\overline{s}_{k+1} - \lambda \overline{d}_{k+1} >\overline{s}_{k} - \lambda \overline{d}_k$ when $k \leq n$ and $\lambda > \lambda(s_n)$. As $\lambda(s_n) \geq \lambda(s_k)$ by Lemma \ref{lem:increasing_whittle_index}, we get $\lambda > \lambda(s_k)$. Hence, according to \eqref{eq:L2}, $\overline{s}_k - \lambda \overline{d}_k < \overline{s}_{k+1} - \lambda \overline{d}_{k+1}$. Thus, $\overline{s}_k - \lambda \overline{d}_k$ is strictly increasing with respect to $k$ when $k \leq n$ and $\lambda > \lambda(s_n)$, which implies that $\overline{s}_{n+1} - \lambda \overline{d}_{n+1} > \overline{s}_{k} - \lambda \overline{d}_k$ for all $k \leq n$.  

Finally, to prove the last condition, we show that $\overline{s}_{k} - \lambda \overline{d}_k$ is decreasing with respect to $k$ when $k > n$ and $\lambda \leq \lambda(s_{n+1})$. As $\lambda(s_{n+1}) \leq \lambda(s_k)$ by Lemma \ref{lem:increasing_whittle_index}, we get $\lambda \leq \lambda(s_k)$. Hence, according to \eqref{eq:L1}, $\overline{s}_k - \lambda \overline{d}_k \ge \overline{s}_{k+1} - \lambda \overline{d}_{k+1}$. Thus, $\overline{s}_k - \lambda \overline{d}_k$ is decreasing with respect to $k$ when $k > n$ and $\lambda \leq \lambda(s_{n+1})$, which implies that $\overline{s}_{n+1} - \lambda \overline{d}_{n+1} \geq \overline{s}_{k} - \lambda \overline{d}_k$ for all $k > n$.
\end{IEEEproof}

The next theorem provides the optimal threshold value as a function of $\lambda$. 
\begin{theorem}\label{theo:expression_threshold_policy_in_function_lambda}
The optimal threshold value satisfies the following conditions:
\begin{enumerate}[label=(\roman*)]
\item If $\lambda \leq \lambda(s_0)$, then the optimal threshold value is equal to $s_0$,
\item If $\lambda(s_n) < \lambda \leq \lambda(s_{n+1})$, then the optimal threshold  value is equal to $s_{n+1}$.
\item If $\lambda \ge \lim_{k \rightarrow \infty} \lambda(s_{k})$, then the optimal threshold value is infinite.
\end{enumerate}
\end{theorem}
\begin{IEEEproof}
When $\lambda \leq \lambda(s_0)$, we have $\lambda \leq \lambda(s_k)$ for all $k \geq 0$. Hence, from~\eqref{eq:L1}, we get $\overline{s}_k - \lambda \overline{d}^k \ge \overline{s}_{k+1} - \lambda \overline{d}_{k+1}$. Hence, in this regime for $\lambda$, we can deduce that the optimal threshold value is $s_0$, as $\overline{s}_0 - \lambda \overline{d}_0 \ge \overline{s}_{k+1} - \lambda \overline{d}_{k+1}$. That completes the proof of the first statement.

However, when $\lambda(n)< \lambda \leq \lambda(n+1)$, from \eqref{eq:L3} and \eqref{eq:L4}, we obtain that $\overline{s}_{n+1} - \lambda \overline{d}_{n+1} \geq \overline{s}_{k} - \lambda \overline{d}_k$, for all $k \geq 0$. Therefore, in this regime for $\lambda$, we can deduce that the optimal threshold value is $s_{n+1}$. That completes the proof of the second statement. 

Let $\lim_{k \to \infty} \lambda(s_k) =\lambda_{\infty}$. Since $\lambda(.)$ is increasing in $s_k$, then $\lambda(s_k) < \lambda_{\infty}$ for all $k$. Thus, when $\lambda_{\infty} \le \lambda$, from~\eqref{eq:L2}, we have $\overline{s}_k - \lambda \overline{d}_k < \overline{s}_{k+1} - \lambda \overline{d}_{k+1}$ as $\lambda(s_k) < \lambda_{\infty} \le \lambda$ for all $k$. Note that the optimal threshold $s_n$ cannot be finite, otherwise $\overline{s}_n - \lambda \overline{d}_n \ge \overline{s}_{n+1} - \lambda \overline{d}_{n+1}$, which is a contradiction. That completes the proof of the third statement.
\end{IEEEproof}

\begin{remark}
Theorem~\ref{theo:expression_threshold_policy_in_function_lambda} delineates three distinct regimes for \( \lambda \). Recall that $\lambda$ is a design parameter, representing an additional positive cost incurred due to the energy consumed for jamming. Depending on the regime in which \( \lambda \) falls into, the optimal threshold value can take one of three possible forms. In the first regime, the threshold is \( s_0 \), indicating that jamming should be performed continuously. In the second regime, the threshold is \( s_{n+1} \), indicating that jamming should be deferred until the EAoII becomes large enough. In the third regime, the threshold becomes \( \infty \), signifying that no jamming should be performed at all.
\end{remark}

\section{Multi-Source Scenario}\label{sec:multi-source}
In this section, we extend our results in Section II to a multi-source scenario in which we deal with multiple sources that send their status updates to a remote monitor over orthogonal unreliable channels, and an adversary that can strategically jam a subset of the channels. The model of each subsystem, comprising a source, a channel, and the monitor, are as before; the adversary acts as an agent; and the performance of status updating for each subsystem is measured based on the associated AoII at the monitor. We first formulate the problem as a RMAB, and then develop a well-performing heuristic jamming policy based on the Whittle's index.

\subsection{Problem Statement}\label{sec:prob_form}
Let $N$ be the total number of subsystems, $p_i$ be the probability that the packet sent by the source $i$ is successfully decoded, $r_i$ be the probability that the source $i$ transitions from one state to another, and $q_i$ be the probability that the adversary succeeds in jamming the channel $i$. Due to the energy constraint, the adversary can jam only $M < N$ channels. A jamming policy $\phi$ is defined as a sequence of actions $\phi=(\boldsymbol{d}_{\phi}(0),\boldsymbol{d}_{\phi}(1),\ldots)$, where $\boldsymbol{d}_{\phi}(t)=(d^1_{\phi}(t),d^2_{\phi}(t),\ldots,d^{N}_{\phi}(t))$ is a binary vector such that $d^i_{\phi}(t)=1$ if channel $i$ is jammed at time $t$. Let $\boldsymbol{s}(t)=(s^{1}(t),\ldots,s^{N}(t))$ be a vector at time $t$ such that $s^i(t)$ is the EAoII of subsystem $i$ at time $t$. Our aim is to find the optimal jamming policy that maximizes the total average EAoII subject to the constraint on the number of channels under attack. This can be formulated~as
\begin{subequations}\label{eq:original_problem}
\begin{align}
& \underset{\phi\in \Phi}{\text{maximize}} \ \lim_{T\to\infty} \text{inf}\:\frac{1}{T}\mathbb{E}^{\phi\in \Phi}\Big[\sum_{t=0}^{T-1}\sum_{i=1}^{N}s^i_{\phi}(t) \big|\boldsymbol{s}(0)\Big],\\
& \text{subject to} \ \sum_{i=1}^{N}d^{i}_{\phi}(t)\leq\alpha N \quad \forall t \ge 0
\end{align}
\end{subequations}
where $\Phi$ denotes the set of all causal jamming policies and $\alpha=M/N$.

\subsection{Lagrangian Relaxation}
The optimization problem in \eqref{eq:original_problem} can be seen as a RMAB, and hence is PSPACE-Hard~\cite{papadimitriou1999complexity}. To mitigate this difficulty, we will adopt a well-established heuristic solution, known as the Whittle's index policy~\cite{weber1990index}. The key for defining the Whittle's index policy is the Lagrangian relaxation, which consists of relaxing the constraint on the available resources by letting it be satisfied on average rather than at each time. For our problem, the relaxed optimization problem is expressed as
\begin{subequations}\label{eq:relaxed_problem}
\begin{align}
& \underset{\phi\in \Phi}{\text{maximize}} \ \lim_{T\to\infty} \text{inf}\:\frac{1}{T}\mathbb{E}^{\phi}\Big[\sum_{t=0}^{T-1}\sum_{i=1}^{N}s^i_{\phi}(t) \big|\boldsymbol{s}(0)\Big],\\
& \text{subject to} \ \lim_{T\to\infty} \text{inf} \: \frac{1}{T}\mathbb{E}^{\phi}\Big[\sum_{t=0}^{T-1}\sum_{i=1}^{N}d^{i}_{\phi}(t)\Big] \leq \alpha N .
\end{align}
\end{subequations}

The Lagrangian function $f(W,\phi)$ associated with the problem~\eqref{eq:relaxed_problem} is defined as
\begin{align}\label{eq:lag-func}
f(W,\phi) &= \!\lim_{T\to\infty} \text{inf} \: \frac{1}{T}\mathbb{E}^{\phi}\Big[\sum_{t=0}^{T-1}\sum_{i=1}^{N}s^i_{\phi}(t) \!-\! Wd^{i}_{\phi}(t) \big|\boldsymbol{s}(0)\Big] \nonumber\\
&\ \qquad \qquad +W\alpha N
\end{align}
where $W \geq 0$ is a penalty for jamming channels. Therefore, our next objective is to solve the following optimization problem:
\begin{align}\label{eq:dual_problem}
\underset{\phi\in \Phi}{\text{maximize}} \ f(W,\phi) .
\end{align}
As the term $W\alpha N$ in \eqref{eq:lag-func} is independent of $\phi$, it can be discarded in the optimization problem. Bearing that in mind, we seek to obtain the Whittle's index jamming policy. We first narrow down our focus to the one-dimensional version of the problem in \eqref{eq:dual_problem}. One can show that the $N$-dimensional problem can be decomposed into $N$ separate one-dimensional problems, each of which can be tackled independently~\cite{kriouile2024asymptotically}. The one-dimensional problem for subsystem~$i$~is
\begin{align}\label{eq:individual_dual_problem}
\underset{\phi\in \Phi}{\text{maximize}} \ \lim_{T\to\infty} \text{inf}\:\frac{1}{T}\mathbb{E}^{\phi}\Big[ \sum_{t=0}^{T-1}s^i_{\phi}(t) - W d^i_{\phi}(t) \big| s^i(0)\Big] .
\end{align}
Now, we need to specify the structure of the optimal solution to this one-dimensional problem. Note that, by replacing $W$ with $\lambda$ in the problem in~\eqref{eq:individual_dual_problem}, we end up with the same problem as in~\eqref{eq:closed_form_express}. Thus, the optimal solution to the problem in~\eqref{eq:individual_dual_problem} is given by Theorem~\ref{theo:threshold_policy}: the optimal policy for a given Lagrangian parameter $W$ is a threshold jamming policy.

\subsection{Whittle's Index Jamming Policy}
Next, we study the existence of the Whittle's indices. Let $s_n^i$ be the state of subsystem $i$ satisfying $s_n^i \in \mathbb{S}^i$, where
\begin{align}
\mathbb{S}^i = &\Big \{s_n^i \big| s_n^i=\frac{1}{2r_i}\big [1+(1-2r_i)^{n+1} \nonumber\\[1.5\jot]
&\qquad \qquad \qquad -2(1-r_i)^{n+1} \big ] , n \ge 0 \Big \}
\end{align} 
and $d_n^i(W)$ be the action at state $s_n^i$ associated with subsystem $i$ under the optimal threshold jamming policy given the Lagrangian parameter $W$. We first consider the steady-state form of the problem in \eqref{eq:individual_dual_problem} under the threshold jamming policy with parameter $s^i_n$, i.e.,
\begin{align}\label{thresholdobjective}
& \underset{n\in \mathbb{N}}{\text{maximize}} \ \overline{s}_{n}^i - W \overline{d}_n^i
\end{align}
where $\overline{s}_{n}^i$ and $\overline{d}_n^i$ are given in \eqref{eq:ss-EAoII} and \eqref{eq:ss-AAT}, respectively, when replacing $p$, $q$, and $r$ with $p_i$, $q_i$, $r_i$. To ensure the existence of the Whittle's indices, it is imperative to establish the indexability property for all subsystems. To accomplish this, we initially formalize the concepts of indexability and the Whittle's index in the following definitions.
\begin{definition}
In the context of the problem in~\eqref{eq:individual_dual_problem}, for a given $W$, we define $D^i(W)$ as the set of states in which the optimal action for each subsystem $i$ is passive, i.e., $d^i_n(W) = 0$. In other words, $s^i_{n} \in D^i(W)$ if and only if the optimal action at state $s^i_{n}$ is passive.
\end{definition}
Note that $D^i(W)$ is well defined as the optimal solution of the problem in~\eqref{eq:individual_dual_problem} is a stationary policy, more precisely, a threshold jamming policy.

\begin{definition}\label{def:Whitt_index}
The problem in~\eqref{thresholdobjective} is indexable if the set of states in which the passive action is the optimal action increases with $W$, i.e., $W' < W \Rightarrow D^i(W') \subseteq D^i(W)$. In this case, the Whittle's index at state $s^i_{n}$ is defined as
\begin{equation}\label{eq:exp-WI}
W(s^i_{n})=\min \Big\{W \big|s^i_{n} \in D^i(W) \Big\}.
\end{equation} 
\end{definition}
\begin{proposition}
For each subsystem $i$, the one-dimensional problem in~\eqref{eq:individual_dual_problem} is indexable.
\end{proposition}
\begin{IEEEproof}
To prove the claim, it is sufficient to show that $\overline{d}^i_n$ is decreasing with respect to $n$. Indeed, we can~write
\begin{align}
&\overline{d}^i_{n+1}-\overline{d}^i_n \nonumber\\
&\ =-\frac{(1-p_i)^n p_i(1-q_i)}{[1-q_i+(1-p_i)^{n+1}q_i][1-q_i+(1-p_i)^nq_i]} \nonumber \\
&\ \le 0.
\end{align}
Hence, the problem~\eqref{thresholdobjective} is indexable.
\end{IEEEproof}

Now that the indexability property has been established, we can assert the existence of the Whittle's index. Building on the procedure introduced in \cite[Algorithm 1]{kriouile2024asymptotically}, we propose a tailored algorithm that iteratively computes the Whittle's index values for each subsystem $i$. The following lemma shows that the output of this algorithm is compatible with Definition~\ref{def:Whitt_index}.

\begin{algorithm}
\caption{Whittle Index of Subsystem $i$}\label{euclid}
\begin{algorithmic}[1]
\State Initialization: $j = 0$
\State Find $W_0^i=\underset{n \in \mathbb{N}}{\text{inf}}\frac{\overline{s}_n^i-\overline{s}_0^i}{\overline{d}_n^i-\overline{d}_0^i}$
\State Define $n_1^i$ as the largest minimizer of the expression in~step~2
\State Let $W(s_k^i)=W_0^i$ for all $k \leq n_1^i$
\While{True}
\State $j=j+1$
\State Define $M_j^i$ as the set $\{0,\cdots,n_{j}^i\}$
\State Find $W_j^i=\underset{n \in \mathbb{N}\setminus M_j^i} {\text{inf}}\frac{\overline{s}_n^i-\overline{s}_{n_j}^i}{\overline{d}_n^i-\overline{d}_{n_j}^i}$
\State Define $n_{j+1}^i$ as the largest minimizer of the expression in step 8
\State Let $W(s_k^i)=W_j^i$ for all $n_{j}^i<k \leq n_{j+1}^i$
\EndWhile
\State Output $W(s_k^i)$ as the Whittle's index at state $s_k^i$
\end{algorithmic}
\end{algorithm}

\begin{lemma}\label{rem:WI_exp_alg}
The expression provided by Algorithm~\ref{euclid} for $W(s_k^i)$ is equivalent to the Whittle's index defined in \eqref{eq:exp-WI}.
\end{lemma}
\begin{IEEEproof}
The proof follows directly the results in \cite[Proposition 4]{kriouile2024asymptotically}.
\end{IEEEproof}

In the next lemma, we provide a useful property that is satisfied by the sequences $\overline{s}_{k}^i$ and $\overline{d}_{k}^i$.

\begin{lemma} \label{lem:th_triangulaire}
For each subsystem $i$, if $\overline{d}_k^i$ is decreasing with $k$ and
\begin{align}\label{eq:lemma6-hyp}
 \frac{\overline{s}_{k+1}^i-\overline{s}_{k}^i}{\overline{d}_{k+1}^i-\overline{d}_{k}^i} < \frac{\overline{s}_{k+2}^i-\overline{s}_{k+1}^i}{\overline{d}_{k+2}^i-\overline{d}_{k+1}^i}   
\end{align}
for all $k \ge 0$, then
\begin{align}\label{eq:ineq_triang}
\frac{\overline{s}_n^i-\overline{s}_{k}^i}{\overline{d}_n^i-\overline{d}_{k}^i} \ge \frac{\overline{s}_{k+1}^i-\overline{s}_{k}^i}{\overline{d}_{k+1}^i-\overline{d}_{k}^i}
\end{align}
for any $n>k$.
\end{lemma}
\begin{IEEEproof}
We fix $k \ge 0$ and prove the result by induction. For $n=k+1$, we can write
\begin{align}\label{eq:lemma6-A}
\frac{\overline{s}_n^i-\overline{s}_{k}^i}{\overline{d}_n^i-\overline{d}_{k}^i} = \frac{\overline{s}_{k+1}^i-\overline{s}_{k}^i}{\overline{d}_{k+1}^i-\overline{d}_{k}^i}   \ge \frac{\overline{s}_{k+1}^i-\overline{s}_{k}^i}{\overline{d}_{k+1}^i-\overline{d}_{k}^i}
\end{align}
where the strict inequality comes from the hypothesis of the lemma. Therefore, we have
\begin{align}
\frac{a_{k+1}-a_{k-1}}{b_{k+1}-b_{k-1}}& > \frac{a_k-a_{k-1}}{b_k-b_{k-1}} \Big(\frac{b_{k+1}-b_k}{b_{k+1}-b_{k-1}}+\frac{b_k-b_{k-1}}{b_{k+1}-b_{k-1}}\Big) \nonumber\\[2\jot]
&=\frac{a_k-a_{k-1}}{b_k-b_{k-1}}.
\end{align}
Now, we assume that \eqref{eq:ineq_triang} is true for a certain $n$ strictly higher than $k$, and show that \eqref{eq:ineq_triang} also hold for $n+1$:
\begin{align}
\frac{\overline{s}_{n+1}^i-\overline{s}_{k}^i}{\overline{d}_{n+1}^i-\overline{d}_{k}^i} 
=& \frac{\overline{s}_{n+1}^i-\overline{s}_{k}^i+\overline{s}_{n}^i-\overline{s}_{n}^i}{\overline{d}_{n+1}^i-\overline{d}_{k}^i}\nonumber\\[2\jot]
=& \frac{\overline{s}_{n+1}^i-\overline{s}_{n}^i}{\overline{d}_{n+1}^i-\overline{d}_{k}^i}+\frac{\overline{s}_{n}^i-\overline{s}_{k}^i}{\overline{d}_{n+1}^i-\overline{d}_{k}^i} \nonumber
\end{align}
\begin{align}
\ge &  \underbrace{\frac{(\overline{s}_{k+1}^i-\overline{s}_{k}^i)(\overline{d}_{n+1}^i-\overline{d}_{n}^i)}{(\overline{d}_{k+1}^i-\overline{d}_{k}^i)(\overline{d}_{n+1}^i-\overline{d}_{k}^i)}}_{(a)}+\underbrace{\frac{(\overline{s}_{k+1}^i-\overline{s}_{k}^i)(\overline{d}_{n}^i-\overline{d}_{k}^i)}{(\overline{d}_{k+1}^i-\overline{d}_{k}^i)(\overline{d}_{n+1}^i-\overline{d}_{k}^i)}}_{(b)} \nonumber\\[1\jot]
=&\frac{\overline{s}_{k+1}^i-\overline{s}_{k}^i}{\overline{d}_{k+1}^i-\overline{d}_{k}^i} \Big [ \frac{\overline{d}_{n+1}^i-\overline{d}_{n}^i}{\overline{d}_{n+1}^i-\overline{d}_{k}^i}+\frac{\overline{d}_{n}^i-\overline{d}_{k}^i}{\overline{d}_{n+1}^i-\overline{d}_{k}^i} \Big ] \nonumber \\[2\jot]
=&\frac{\overline{s}_{k+1}^i-\overline{s}_{k}^i}{\overline{d}_{k+1}^i-\overline{d}_{k}^i}
\end{align}
where $(a)$ comes from \eqref{eq:lemma6-hyp} and the fact that $\overline{d}_k^i$ is strictly decreasing with respect to $k$, and (b) comes from \eqref{eq:lemma6-A} and the fact that $\overline{d}_k^i$ is decreasing with respect to $k$. 
Therefore, the inequality also holds for $n+1$. This concludes the proof. 
\end{IEEEproof}


Finally, in the next theorem, we derive a closed-form expression of the Whittle's index corresponding to each subsystem $i$, which enables us to propose the Whittle index jamming policy as a solution to the original problem in~\eqref{eq:original_problem}.
\begin{theorem}\label{theo:Whittle_index_expressions}
For each subsystem $i$ and state $s_n^{i}$, the Whittle's index is obtained as
\begin{align}
W(s_n^i)=\lambda_i(s_n)=\frac{\overline{s}_{n+1}^i-\overline{s}_{n}^i}{\overline{d}_{n+1}^i-\overline{d}_{n}^i}
\end{align}
where $\lambda_i(.)$ is given by the following expression, when replacing $p$, $q$, and $r$ with $p_i$, $q_i$, $r_i$:
\begin{align}\label{eq:exp_lambda}
\lambda(s_n) &= \frac{p q (1-r)}{r(1-z)}-\frac{pq(1-2r)}{2r(1-y)} \nonumber\\[2\jot]
&\ \ -\frac{(1-r)pq(1-q)(1-r)^{n+1}(1-z)}{r\big (1-w\big)\big(1-z\big )} \nonumber\\[2\jot]
&\ \ -\frac{(1-r)pq(1-p)^{n+1}q(1-r)^{n+1}r}{r\big (1-w\big)\big(1-z\big)} \nonumber\\[2\jot]
&\ \ +\frac{(1-2r)pq (1-q)(1-2r)^{n+1}(1-y)}{2r\big(1-x\big)\big(1-y\big)} \nonumber\\[2\jot]
&\ \ +\frac{(1-2r) 2pq (1-p)^{n+1}q(1-2r)^{n+1}r}{2r\big(1-x\big)\big(1-y\big)}
\end{align}
where $w = (1-r)(1-p+pq)$, $x = (1-2r)(1-p+pq)$, $y = (1-2r)(1-p)$, and $z = (1-r)(1-p)$.
\end{theorem}
\begin{IEEEproof}
According to Lemma~\ref{rem:WI_exp_alg}, we need to show that $\lambda_i(.)$ satisfies $W_j^i = \lambda_i(s_n)$ for all $n_{j}^i<n \le n_{j+1}^i$, where the sequences $n_j^i$ and $W_j^i$ are provided by Algorithm~\ref{euclid}. Let the index $j$ denotes $j$th iteration step. We first show that
\begin{align}
\frac{\overline{s}_n^i-\overline{s}_{j}^i}{\overline{d}_n^i-\overline{d}_{j}^i} \ge \frac{\overline{s}_{j+1}^i-\overline{s}_{j}^i}{\overline{d}_{j+1}^i-\overline{d}_{j}^i}
\end{align}
for all $n>j$. By Lemma \ref{lem:increasing_whittle_index}, we know that
\begin{align}
\frac{\overline{s}_{k+1}^i-\overline{s}_{k}^i}{\overline{d}_{k+1}^i-\overline{d}_{k}^i} < \frac{\overline{s}_{k+2}^i-\overline{s}_{k+1}^i}{\overline{d}_{k+2}^i-\overline{d}_{k+1}^i}  
\end{align}
for all $k \ge 0$. Hence, by Lemma \ref{lem:th_triangulaire}, we can write that
\begin{align}
\frac{\overline{s}_n^i-\overline{s}_{j}^i}{\overline{d}_n^i-\overline{d}_{j}^i} \ge \frac{\overline{s}_{j+1}^i-\overline{s}_{j}^i}{\overline{d}_{j+1}^i-\overline{d}_{j}^i}   
\end{align}
for all $n > j$. Thus, the minimizer of $(\overline{s}_n^i-\overline{s}_{j}^i)/(\overline{d}_n^i-\overline{d}_{j}^i)$ over $n$ at step $j$ is $j+1$. As a result, the Whittle's index at state $s_j^i$ is obtained as
\begin{align}
W(s_j^i)=W_j^i=\frac{\overline{s}_{j+1}^i-\overline{s}_{j}^i}{\overline{d}_{j+1}^i-\overline{d}_{j}^i}=\lambda_i(s_j).
\end{align}
This completes the proof.
\end{IEEEproof}

\begin{remark}
The result of Theorem \ref{theo:Whittle_index_expressions} allows the adversary to prioritize which channels to jam based on real-time information, making it highly suitable for dynamic environments. This policy provides a practical and near-optimal solution to the jamming problem in multi-source networks. Note that it has been shown in the literature \cite{kriouile2024asymptotically, larranaga2017asymptotically, kriouile2021global} that the Whittle index policy can achieve asymptotic optimality, meaning that as the number of subsystems grows large, the performance of the Whittle index policy approaches that of the globally optimal solution.
\end{remark}

Accordingly, we can provide the following algorithm for implementation of the Whittle's index jamming policy.
\begin{algorithm}
\caption{Whittle's Index Jamming Policy}
\begin{algorithmic}[1]
\State At each time $t$, compute the Whittle's index of all sources using the expressions given in Theorem~\ref{theo:Whittle_index_expressions}.
\State Jam the $M$ channels corresponding to the $M$ subsystems with the highest Whittle's index values at time $t$.
\end{algorithmic}
\end{algorithm}

\section{Numerical Results}\label{sec:num_reslt}
In this section, we provide numerical results for both single and multi-source scenarios, which complement our theoretical results in Sections II and III. In each scenario, we compare the performance of our proposed jamming policy with that of the uniform random jamming policy.

\subsection{Single-Source Scenario}
For the single-source scenario, we showcase the performance of our optimal jamming policy, and compare that with performance of the uniform random jamming policy, which is the one that jams the channel with probability $0.5$ at every time. Recall that $\lambda$ represents the energy cost, or equivalently the amount of the energy consumed by the adversary when it decides to jam the channel. 

\begin{figure}[t!]
\centering
\includegraphics[width=0.5\textwidth,trim={6mm 0 2mm 8mm}]{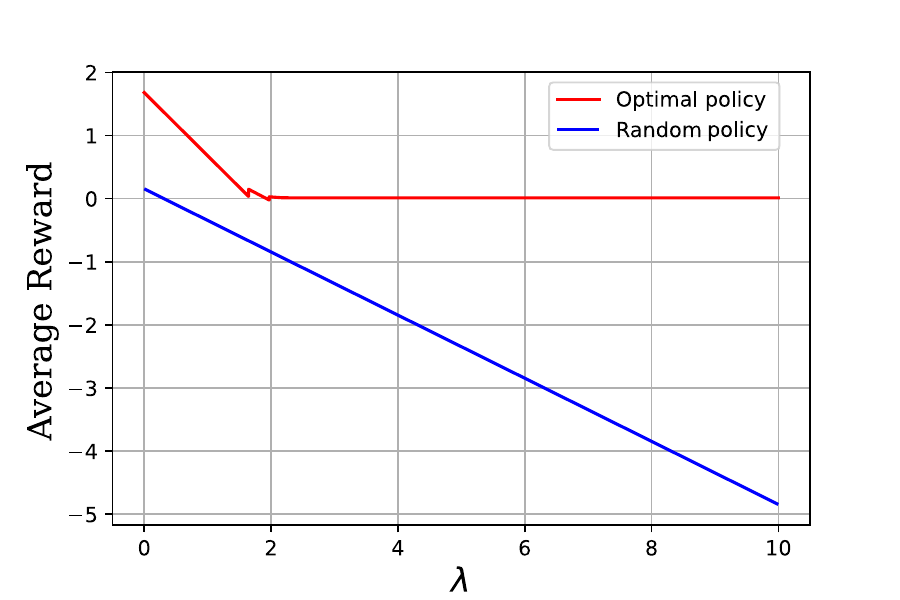}
\caption{Comparison between the optimal policy and a random policy in terms of the average reward.}
\label{fig:comp_reward_first_sc_single_src}
\end{figure}

Fig.~\ref{fig:comp_reward_first_sc_single_src} illustrates the evolution of the average reward under the optimal jamming policy derived in Theorem~\ref{theo:expression_threshold_policy_in_function_lambda} and under the uniform random jamming policy as a function of $\lambda$. We supposed that $\lambda$ varies from $0$ to $10$ with step size $0.001$, and we considered the following parameters $p=q=0.9$ and $r=0.1$. As $\lambda$ increases, the average reward for the optimal solution declines and eventually stabilizes at a constant value. This is because higher values of $\lambda$ result in a greater energy cost for jamming, which in turn diminishes the performance of the optimal strategy.. We also observe that the uniform random jamming policy is sub-optimal, which corroborates our theoretical results.

\begin{figure}[t!]
\centering
\includegraphics[width=0.5\textwidth,trim={6mm 0 2mm 8mm}]{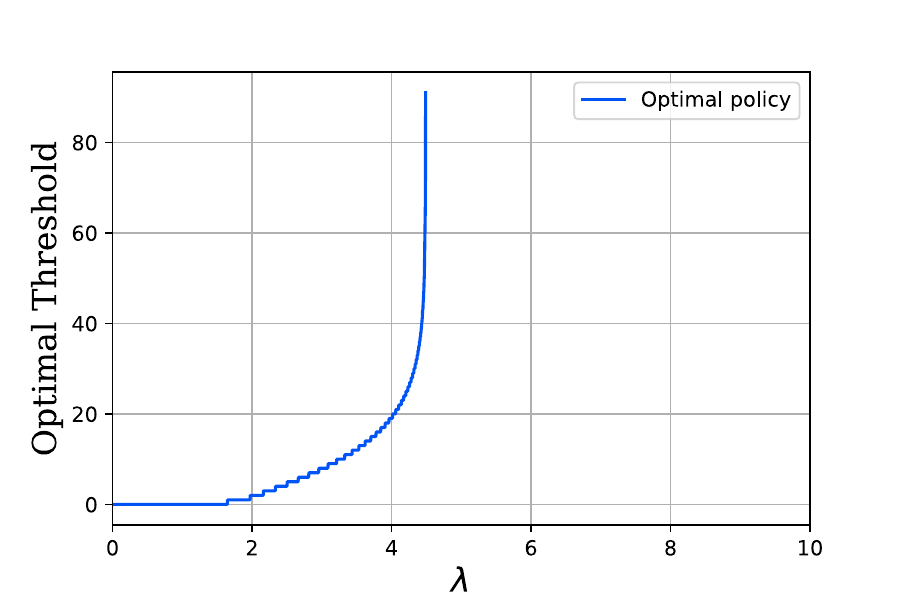}
\caption{Threshold value of the optimal jamming policy as a function of $\lambda$}
\label{fig:threshold_first_sc}
\end{figure}

In addition, Fig.~\ref{fig:threshold_first_sc} illustrates the evolution of the optimal threshold value as a function of $\lambda$. We can observe that the optimal threshold increases exponentially with $\lambda$, and tends to $\infty$ when $\lambda$ is approximately equal to $4.5$. This phenomenon occurs because, as $\lambda$ grows, the adversary consumes more energy for jamming the channel. To compensate for this increase in energy consumption, the adversary reduces the average active time by increasing the threshold. This observation also aligns with Fig.~\ref{fig:comp_reward_first_sc_single_src}, as when $\lambda$ grows the optimal threshold values tends to $\infty$, which implies that the corresponding reward is converging according to Propositions~\ref{prop:average_reward_expression} and \ref{prop:average_active_time}. Moreover, unlike the AoII-based optimal policy derived in our previous work~\cite{kriouile2024optimal}, the threshold here becomes infinite when $\lambda$ exceeds a certain value. This can be explained by the fact that the increasing rate of the average EAoII is of the same order as that of the average AAT in the POMDP. In other words, when $\lambda$ surpasses a certain threshold, regardless of how frequently the adversary targets the channel, the EAoII can never compensate for the average energy consumed if $\lambda$ is sufficiently high. Mathematically, according to Theorem \ref{theo:expression_threshold_policy_in_function_lambda}, the threshold $\lambda$ beyond which attacking the channel becomes inefficient is given by $\frac{pq (1-r)}{r(1-(1-r)(1-p))}-\frac{pq(1-2r)}{2r(1-(1-2r)(1-p))}$.

\begin{figure}[t!]
\centering
\includegraphics[width=0.5\textwidth,trim={6mm 0 2mm 8mm}]{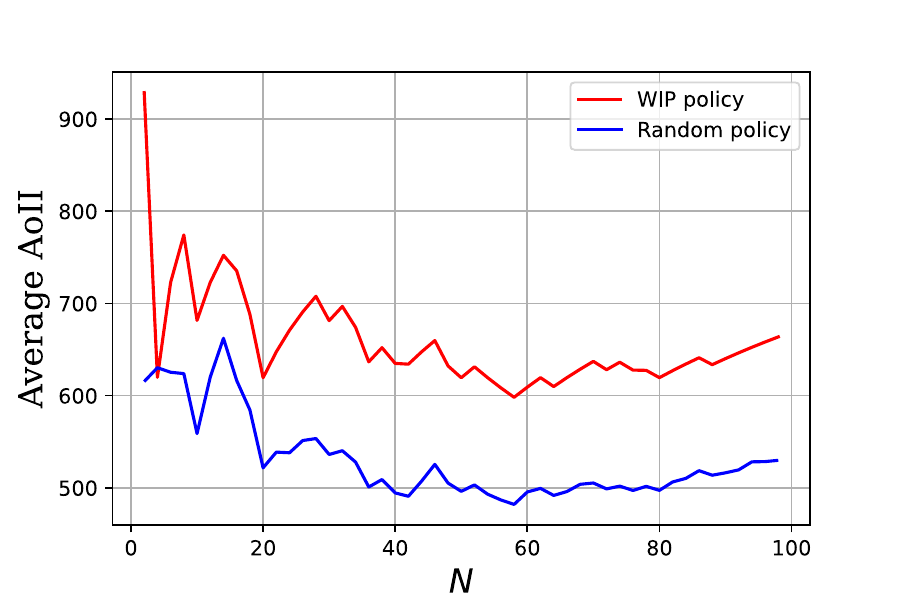}
\caption{Comparison between the Whittle index policy (WIP) and a random policy in terms of the average AoII.}
\label{fig:WIP_vs_RD_policy}
\end{figure}

\subsection{Multi-Source Scenario}
For the multi-source scenario, we showcase the performance of our Whittle index jamming policy, and compare that with the performance of the random policy in terms of the total average AoII. The random policy selects $M$ channels randomly among the $N$ channels in the system at each time $t$. For that, we consider two different classes. The sources' parameters in class one are $p_1=0.2$, $q_1=0.2$, $r_1=0.4$, and in class two are $p_2=0.8$, $q_2=0.8$, $r_2=0.2$. We assume that the proportion of sources in each class with respect to the total number of sources in the system is $0.5$, and that the adversary can jam at most half of the channels $M=N/2$, where $N$ represents the total number of the sources. Fig.~\ref{fig:WIP_vs_RD_policy} illustrates the evolution of the average AoII under the Whittle index jamming policy and under the random jamming policy as a function of $N$. We can notice that our policy outperforms the random policy when~$N$~grows.

\section{Conclusion}\label{sec:conclusion}
In this paper, we studied the problem of denial-of-service attacks against status updating in a partially observable environment. The target system was modeled by Markov chains and i.i.d. channels, and the performance of status updating was measured based on the AoII. We derived the structures of the optimal jamming policies in single and multi-source scenarios. For the single-source case, we demonstrated that the optimal jamming strategy follows a threshold policy and designed a low-complexity algorithm to determine the optimal threshold. Additionally, for the multi-source scenario, we adopted the Lagrangian relaxation approach to develop a low-complexity and well-performing policy, dubbed Whittle index jamming policy. Finally, our numerical results showed that the proposed policy for both scenarios can outperform random policies, corroborating thus our theoretical findings.

\bibliography{DOS.bib}
\bibliographystyle{ieeetr}

\vspace{-100mm}

\end{document}